\documentclass[conference]{IEEEtran}
\ifCLASSINFOpdf
 \usepackage[pdftex]{graphicx}
\else
\fi
\ifCLASSOPTIONcompsoc
 \usepackage[caption=false,font=normalsize,labelfont=sf,textfont=sf]{subfig}
\else
 \usepackage[caption=false,font=footnotesize]{subfig}
\fi

\usepackage{xcolor}

\begin{document}
%
\title{Towards human-agent knowledge fusion (HAKF) in support of distributed coalition teams}

\author{
\IEEEauthorblockN{
Dave Braines\IEEEauthorrefmark{1}, 
Federico Cerutti\IEEEauthorrefmark{2}\IEEEauthorrefmark{3},
Marc Roig Vilamala\IEEEauthorrefmark{3},
Mani Srivastava\IEEEauthorrefmark{4},\\
Lance Kaplan\IEEEauthorrefmark{5}
Alun Preece\IEEEauthorrefmark{3}, 
Gavin Pearson\IEEEauthorrefmark{6}\\\\
}

\IEEEauthorblockA{
\IEEEauthorrefmark{1}IBM Research Europe, Hursley Park, Hampshire, UK. SO21 2JN\\
\IEEEauthorrefmark{2}Department of Information Engineering, University of Brescia, 25123 Brescia, Italy,\\
\IEEEauthorrefmark{3}Crime and Security Research Institute, Cardiff University,\\
\IEEEauthorrefmark{4}University of California, Los Angeles, USA,\\
\IEEEauthorrefmark{5}US Army Research Laboratory, Adelphi, MD 20783, USA,\\
\IEEEauthorrefmark{6}Defence Science and Technolog Laboratory, Porton Down, Salisbury, Wiltshire, UK. SP4 0JQ
}
}


\maketitle

\begin{abstract}
Future coalition operations can be substantially augmented through agile teaming between human and machine agents, but in a coalition context these agents may be unfamiliar to the human users and expected to operate in a broad set of scenarios rather than being narrowly defined for particular purposes. In such a setting it is essential that the human agents can rapidly build trust in the machine agents through appropriate transparency of their behaviour, e.g., through explanations. The human agents are also able to bring their local knowledge to the team, observing the situation unfolding and deciding which key information should be communicated to the machine agents to enable them to better account for the particular environment. In this paper we describe the initial steps towards this human-agent knowledge fusion (HAKF) environment through a recap of the key requirements, and an explanation of how these can be fulfilled for an example situation. We show how HAKF has the potential to bring value to both human and machine agents working as part of a distributed coalition team in a complex event processing setting with uncertain sources.

\end{abstract}


%
\IEEEpeerreviewmaketitle

\section{Introduction}
\label{sec:intro}
Our overall goal for human-agent knowledge fusion (HAKF) is to enable rapid exploitation of adaptive coalition situational understanding (CSU) knowledge to inform decision-making across coalitions. This must be realistically achievable in a high-tempo environment that can be applied to a wide variety of domains and operating conditions, and accessed by human users who are not deep technical experts. Specifically we are investigating system architectures to enable demonstrable synergy between machine and human agents seeking to gain actionable insight and foresight in a contested coalition environment.

We have created a rich and flexible graphical environment named Cogni-Sketch\footnote{For more details please refer to https://cogni-sketch.org} (CS) as an experimental embodiment of the HAKF concept, within which complex information can be shared between multiple human and machine agents in the form of a knowledge graph. This simple graph-based representation is extensible to support representation of rich domain-relevant information through the creation or reuse of domain concepts and relationships with well defined semantics (in the form of ontologies). However, these ontologies are not exposed directly to the user but are instead made available as a set of concepts within one or more palettes that appear within the CS environment. These palette items can be used by human or machine agents to represent and link information relevant to the problem-solving task and the wider domain, and can be contributed by any agent at any time. The simple semantics (based on first-order predicate logic) is expressed in the form of inheritance within the concept hierarchy, alongside well-defined named relationships between particular concepts, and the ability to have unnamed concepts and relationships to support customisation and extension during operations.

Internally this visual graph is expressed in a simple JSON-based machine-processable representation that supports querying, graph traversal and other forms of analytics by machine agents, and is accessible in various visual forms (graph, table, timeline, etc) as well as through a conversational interface that enables the human and machine agents to interactively explore (or extend) the graph. This environment is described in more detail in Section \ref{sec:example}.

In addition to agile and flexible human-machine collaboration, another key focus of our research is the ability to span neuro-symbolic processing to bring together the explanatory and instinctive power of more traditional rule-based systems with the predictive power of neural network-based processing, such as machine learning and deep learning systems. These are made available as services within the CS environment, with the ability to connect the inputs or outputs to the conceptual palette items mentioned previously, thereby creating an environment in which such neuro-symbolic alignments can be easily defined.

\begin{figure}[ht]
\includegraphics[width=5cm]{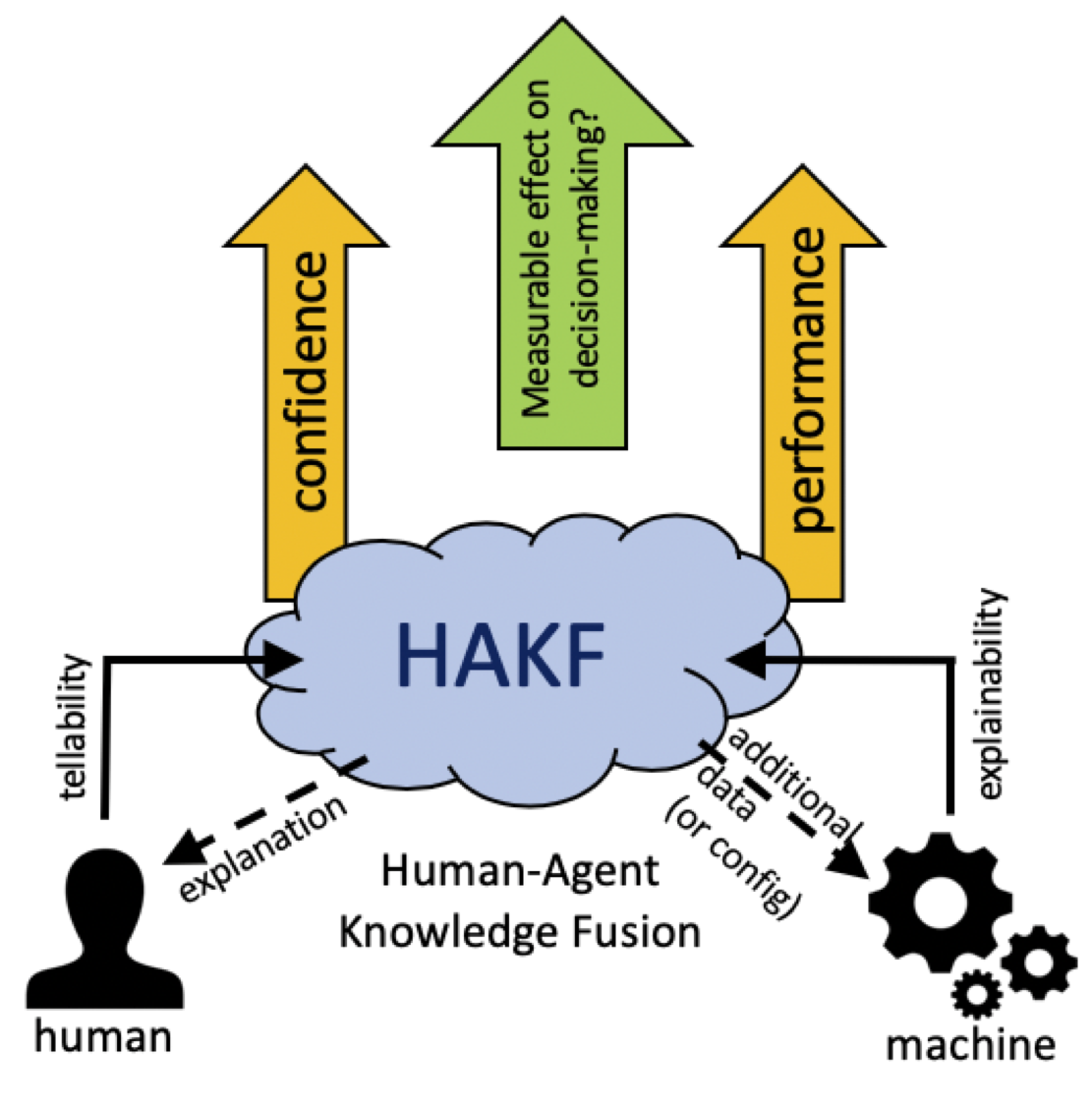}
\centering
\caption{Human-agent knowledge fusion (HAKF) - a mechanism to support complex information exchange between human and machine agents}
\label{fig:hakf}
\end{figure}

Figure \ref{fig:hakf} shows the basic premise behind the HAKF approach: That such an environment must enable collaboration between human and machine agents, specifically through the two communication forms of \emph{tellability} and \emph{explainability} (see following sub-sections for more detail). The overall premise is that HAKF systems with explainability can increase human-agent confidence (through transparency), and systems with tellability can increase machine-agent performance (through customisation and configuration). We believe that hybrid systems with improved confidence and performance can have a measurable effect on decision making, especially in rapidly evolving congested and contested edge-of-network settings where it is infeasible to pre-build high quality custom solutions to support the decision-maker in predictable ways.

Plans for evaluating this hypothesis and testing the impact on confidence, performance and any measurable effect on decision making are under development and will involve the environment described in this paper, and potentially aspects of the particular use cases described here. It is likely to take the form of two separate exercises: one with a large group of users sourced through a suitable online platform such as Amazon MTurk\footnote{See https://www.mturk.com/}, carrying out simple collaborative tasks using the explainability and tellability methods described below, and the second as a smaller and more open ended group exercise with subject matter experts such as open source intelligence analysts. In the first case the measurements will be largely obtained from behavioural analytics through instrumentation embedded in the environment, and in the second case the analysis will be more qualitative, and sourced through a post-exercise survey.

Specific details related to the measures taken, the experimental design, and the role of different sub-groups for comparison will be created in future research, but in this paper we outline the environment and two example use cases that can be supported.

\subsection{Tellability}
\label{sec:tell}
This flow is for the case where new information is conveyed from one of the coalition users to the system, often to impart useful and relevant local knowledge that could improve the performance of the system overall. Depending on the role of the user within the system this information could affect the system at any level. For this paper the focus is on the users tasked with rapidly building or deploying the system. A key focus is on supporting these users in configuring the system to rapidly apply it (or refocus it) on a particular situation. It is unlikely that machine agents (such as machine learning or deep learning components) can be retrained in the short timeframe for the agile operations that are the focus of our research, but through tellability the user may be able to connect the lower-level generic classification outputs of such components to higher-level concepts. It is this conflict between the cost and time-taken to retrain custom models, verses the generic capabilities provided by existing pre-trained models that drives our approach. Clearly a superior technical solution would involve custom trained models for the exact situation and data feeds, but the tempo of the unfolding situation and the inability to predict the exact context in advance means that we cannot have the luxury of custom models and instead look at options and architectures for harnessing generic pre-trained models instead. For example, in the activity recognition capability described in Section \ref{sec:example} this could involve the observation of the service in operation on a new coalition camera feed, and the identification of common background classifications that are occurring during `normal operation' of that feed, and that these may vary according to different conditions such as night vs day, or crowded vs sparsely populated scenes. In other words, one useful capability that is enabled through this environment is the ability to rapidly add relevant contextual information to classifier outputs. The users are able to create new workflows from component parts, or add contextual information to existing workflows to increase their value and relevance to any given operation.

\subsection{Explainability}
\label{sec:explain}
Conversely, explainability provides a greater level of transparency into a conclusion or output from an agent within the HAKF environment. Amongst human agents this is a familiar concept and is often invoked through `why?' questions and appropriate responses. Explanations can be well served through interactive discourse (verbally, textually, or through other means such as interactive visualisations) and for the machine agents this might be through traditional Explainable AI techniques \cite{arrieta2020explainable} such as attention or saliency highlighting, or description of configuration, highlighting relevant operating rules or constraints, or training data, as well as the potential to provide certainty information about any of these, for example how much the input data aligned with training data for that classifier \cite{chakraborty2017interpretability} etc.

In Section \ref{sec:example} we define a simple worked example using the selective audio-visual relevance (SAVR) explainability technique --- derived from the research reported in \cite{taylorVADR} --- to highlight the video and audio aspects of an incoming sensor feed that most influenced a particular classification. 

These SAVR explanations and other similar explainability capabilities will often form part of an interactive discourse between agents within the HAKF environment. For example, the use of explanations such as SAVR to determine system operation in order to refine configuration, or align outputs to wider contextual information. This kind of `conversational interaction` involves both explainability and tellability and is explicitly supported by the HAKF environment through fairly simple capabilities that can be used in open and flexible ways. Using these techniques any appropriate configuration information can be used to rapidly tailor generic capabilities to specific situations, or new data feeds, in a quick and intuitive manner.

To summarise: The technical contribution of this work is to outline the way in which generic services can be composed into situation-specific applications with minimal technical effort by the system designer users. The tempo of this is extremely important since it is not credible to pre-define such specific services to cover all possible situations that may arise. Given the coalition edge-of-network operating context and the often urgent nature of these situations, it is important to create and deploy solutions rapidly, and the platform and worked example in this paper are designed to demonstrate how this can become possible.

\section{Worked example}
\label{sec:example}
Our worked example focuses on the role of the system designer \cite{Tomsett:2018}, or someone tasked with configuring a system to support decision-maker users in a CSU context, but is not focused on the eventual end users of that system. See \cite{BarrettPowellSUE} for a full description of such a system, which is built using the same components as described in this paper but from the perspective of the end user; the decision-maker tasked with maintaining situational understanding in a complex and fast-moving environment. The ability for such systems to be rapidly created in response to unfolding situations is very important, since it is this near-real-time dynamic composition of services and capabilities that makes it plausible that such solutions could be credibly created in real situations in the future. So whilst the focus of \cite{BarrettPowellSUE} is on the end user, and describes the capabilities that they can use, in this paper we instead focus on what techniques can be used by system designer users to credibly build such systems from component services.

So, our system designer user user is tasked with taking broad or generic capabilities such as pre-trained machine learning services and rapidly applying them to a particular coalition setting such as monitoring multiple data feeds for potential events that in aggregate may constitute some kind of risk or threat to the broader operation \cite{xing2019deepcep}. These services must therefore be applied to appropriate data feeds or sensors from a variety of coalition partners in different locations, to issue events or alerts as part of this CSU activity.

In earlier work \cite{taylorVADR} we have defined a technique for providing explanations for the classification decision of a multi-modal (audio and video) activity recognition system. The deep neural network model was trained \cite{taylorVADR} using a generic activity dataset (UCF-101 \cite{soomro2012ucf101}) and is therefore able to output classification decisions for a wide range of human activities that can be sensed through audio-visual means such as CCTV or camera sources. These classifications will be for specific activities that are unlikely to be of particular relevance to the domain problem in which they are applied. This means that one key task for these system designer users is the rapid integration of services such as activity recognition to a variety of data feeds, and the subsequent mapping of the generic output classifications to other, domain-relevant classifications. Due to the use of generic pre-trained models, when applying these services to data feeds the user must identify the classes which are commonly matched during normal conditions. These may vary based on the scene of the data feed as well as potentially contextual factors such as lighting conditions (e.g., night vs day) and crowd density vs sparsity.

\subsection{General event mapping}
\label{sec:events}
\begin{figure}[ht]
\includegraphics[width=7cm]{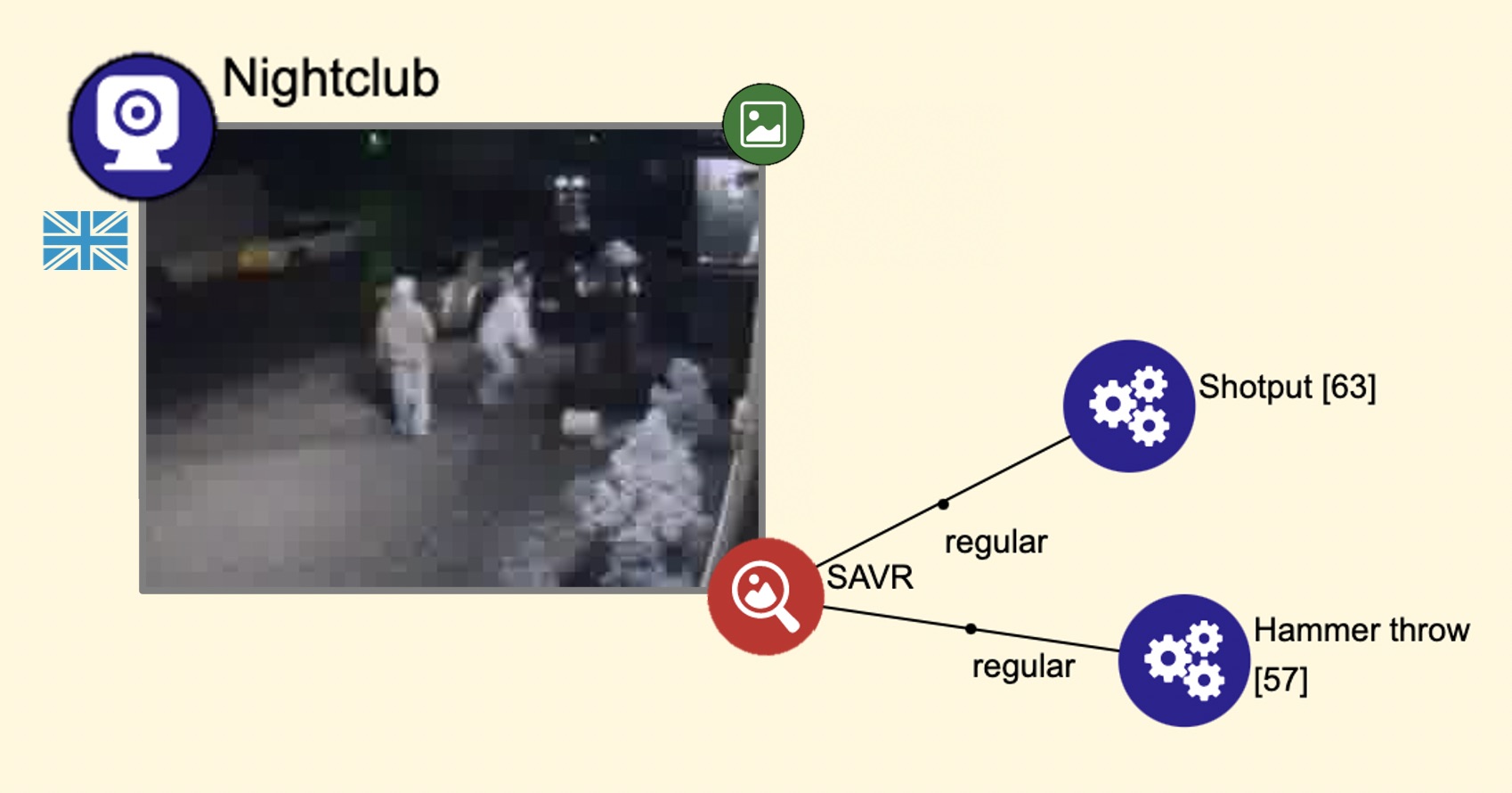}
\centering
\caption{Example customisation of activity recognition analytics (with SAVR explanations) against a coalition data-feed to support situation understanding}
\label{fig:savr-sketch}
\end{figure}

Figure \ref{fig:savr-sketch} shows an example canvas within the Cogni-Sketch environment in which the user has dragged out the video feed from the `Nightclub' camera (A UK asset as indicated by the flag), and has then dropped the `SAVR' explainable analytic service onto this video-audio feed\footnote{For simplicity we refer to the explainable activity recognition service here as `SAVR' though, technically, SAVR refers to the explanation component specifically.}. Immediately upon doing so, any classifications that are predicted by the service are shown as new nodes on the canvas, along with the frequency of classifications for each case. In this example we can see that the activities of `shotput' and `hammer throw' are frequently identified. The user is able to mark these as `regular' classes and therefore enable them to be ignored by any downstream services that may wish to consume the analytic output of the nightclub data feed with SAVR analytics and explanations. Effectively this means that those two classes of `shotput' and `hammer throw' will be ignored by the downstream service(s).

The green icon on the top right enables the SAVR explanations, highlighting the video and audio saliency to the current scene, or historical footage, but these are not described in this paper (refer to \cite{taylorVADR}). A user accessing these explanations, along with associated confidence or certainty metrics, may be seeking visibility of the saliency of parts of the video or audio in order to better understand the way in which activity recognition is being influenced. We are seeking to enable `the user to rapidly calibrate their trust in the system's outputs, spotting flaws in its reasoning or seeing when it is unsure' \cite{tomsett2020rapid}. Such understanding may then inform the user of potentially better configuration choices, and is an example of the explainability-tellability discourse mentioned previously.

\subsection{Defining complex events}
\label{sec:cep}
Another role that the system designer user may need to perform is the definition of complex events in a rapidly unfolding context. The ability to detect a wide variety of simple events from multiple modalities (e.g., video, audio, seismic, textual etc) would already be in place, but the particular way in which these simple events come together to form complex events may vary substantially for each deployment. Contextual factors such as the physical location, the range and location of sensors and the background activities occurring in the scene may all have a bearing on the relevant complex events and the ways in which they may be detected.

\begin{figure}[ht]
\includegraphics[width=8cm]{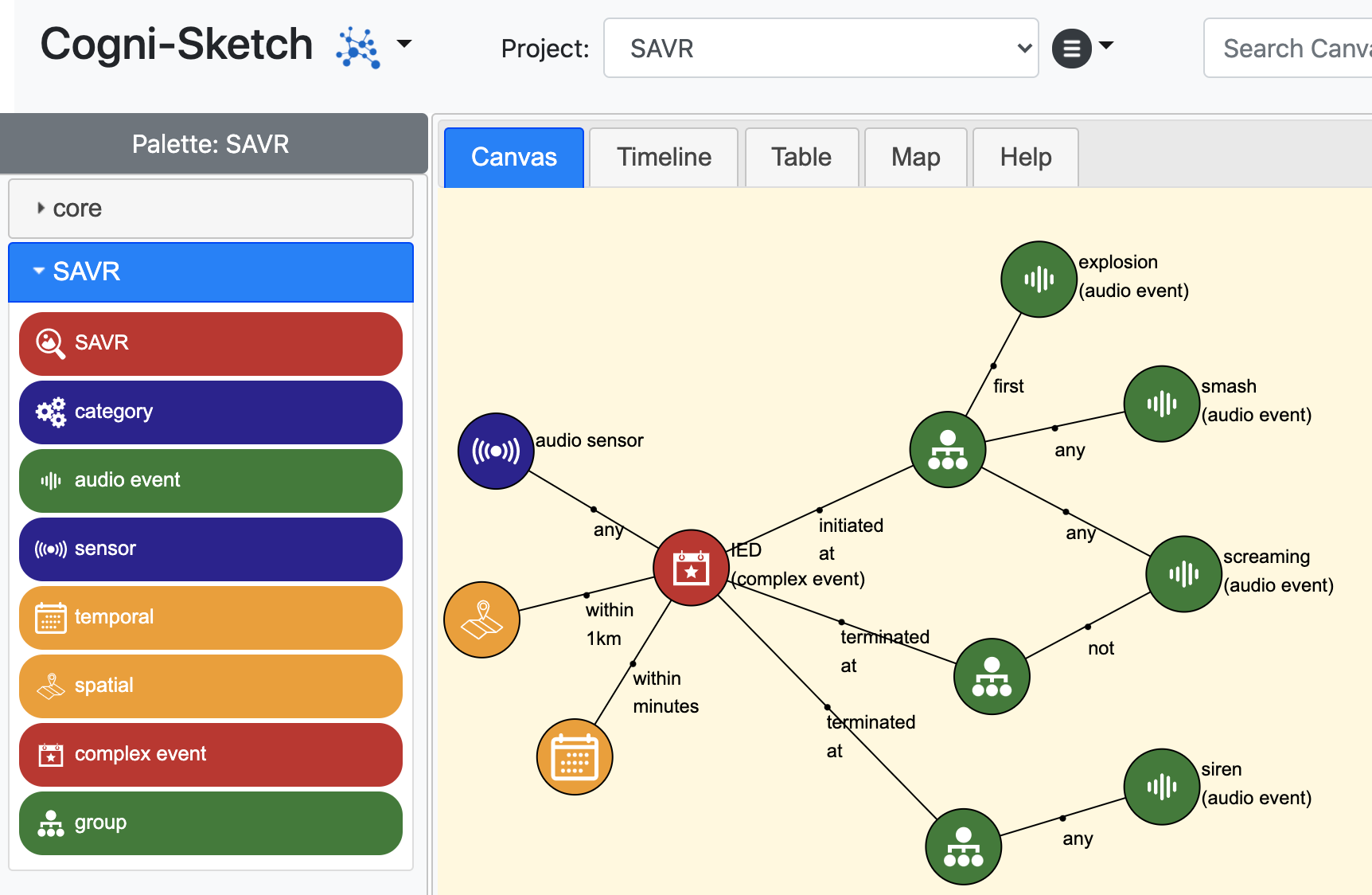}
\centering
\caption{Defining complex events from detectable simple events using audio sensors}
\label{fig:cep}
\end{figure}

In Figure \ref{fig:cep} we show the same Cogni-Sketch environment, but this time the palette is included on the left-hand side of the figure. This is an extensible palette which can be shared between human agents or across operations. In this role we show a palette suitable for defining complex events with spatio-temporal constraints and a pre-defined set of simple (audio) events that are detectable by the system. In this overview paper we include this as a simple example to show the HAKF environment supporting human-agent interaction through tellability and explainability in a second situation. In a subsequent paper we will define the underlying ontology and typology that guides the users through the task of defining complex events.

In the example shown we have defined an `IED' (Improvised Explosive Device) event based on occurrences of semantically relevant simple audio events such as explosion and siren, and have indicated which of these initiate and terminate the complex event, along with simple spatio-temporal constraints. Having composed these definitions the human agent has created knowledge (through tellability) that can be immediately acted upon by machine agents, and if permission is granted this can easily be shared across one or more coalition partners through machine agent communication.

The output of the complex event definition will vary by implementation, but in our demonstration system this generates appropriate fragments of ProbLog which explicitly define these events. Upon detection these higher level complex events can also serve the downstream agents tasked with coalition situational awareness and understanding for this area.

\section{Future work}
\label{sec:future}
We plan to explore techniques for intuitively exposing uncertainty information, especially for multiple dimensions of uncertainty from different sources (both human and machine). We will embed visual techniques for clearly conveying these to the human decision-maker \cite{sensoy2018evidential}. For example, in the context of SAVR explanations, there may be inherent uncertainty associated with the data feed, based on the location, the fidelity, any time-delay, or even the trust associated with the coalition partner that is providing that feed. There may also be inherent certainty in the core classification decision, or the explanation \cite{chakraborty2017interpretability}, as well as potential ambiguity or uncertainty in the alignment of that core classification outcome to any higher-level domain concepts relevant to situation understanding. All of these sources of uncertainty may be potentially important to the decision-maker (or other agents within the system), but how can this be conveyed in a consumable form that does not add substantial cognitive burden? We plan to define various visualisation techniques and asses them in different settings with users, along with a conversational (or visual) interaction to allow the users to explore more deeply when needed.



Finally, this HAKF layer can also be integrated with autonomous agents supporting humans in their activities. Consider, for example, the case where one member of a coalition needs privileged access to some resources or data stream provided by another member. Assuming that such data streams can be guarded by autonomous agents following a set of guidelines the humans provided them with, we can thus embed in our HAKF layer an autonomous agent negotiating with the autonomous guardian(s) of service level agreements involved in accessing the data stream. For example, a request to access a real-time full-definition video stream from a surveillance camera mounted on an unmanned autonomous vehicle (UAV) whose location might be sensitive and cannot be disclosed. In this example the embedded agent would negotiate within the constraints (e.g., a time delay must be imposed), seeking to obtain information relevant to their mission. In previous research we worked on enabling autonomous negotiation in coalition settings \cite{Vente2020IncreasingNP}, and in future work we plan to integrate this capability within the HAKF layer we present in this paper.

\section{Summary}
\label{sec:summary}
In this paper we have described the concept of human-agent knowledge fusion (HAKF) and the initial experimental implementation in the form of Cogni-Sketch (CS). Through a worked example we have shown how this environment can be used to rapidly align data feeds and analytic explainable services (such as SAVR), and align the outputs of these to higher-level concepts that more closely serve the specific operation in each case. We also showed how the same environment can be used to build definitions of complex events rapidly, and in principle we believe the HAKF concept can support many similar activities. The CS environment is inherently extensible through the definition of semantic concepts and relationships which are available to the users via the palette, and the whole knowledge graph is available in a simple JSON format for easy machine agent consumability, and through negotiation can form the basis for coalition interoperability through fine-grained sharing of information from this graph according to policies and agent-managed service level agreements. We show how system designer users are able to use this environment to rapidly define or configure analytic services and complex events for later use in a coalition situational understanding context.

\section*{Acknowledgement}
This research was sponsored by the U.S. Army Research Laboratory and the UK Ministry of Defence under Agreement Number W911NF-16-3-0001. The views and conclusions contained in this document are those of the authors and should not be interpreted as representing the official policies, either expressed or implied, of the U.S. Army Research Laboratory, the U.S. Government, the UK Ministry of Defence or the UK Government. The U.S. and UK Governments are authorized to reproduce and distribute reprints for Government purposes notwithstanding any copyright notation hereon.



\bibliographystyle{IEEEtran}
\bibliography{IEEEabrv, References}
%

%
%
%

\end{document}